\documentclass[pra,twocolumn]{revtex4}
\usepackage{bm,graphicx,amsmath,amssymb}
\usepackage{epsf,graphics,graphicx}
\usepackage{float}
\usepackage{subfigure}
\usepackage{amsthm}
\usepackage{xcolor} 
 
\newcommand{\ket}[1]{|#1\rangle}
\newcommand{\bra}[1]{\langle #1|}

\begin{document}
\title{Geometry along Evolution of Mixed Quantum States}
\author{Erik Sj\"{o}qvist}
\email{erik.sjoqvist@physics.uu.se}
\affiliation{Department of Physics and Astronomy, Uppsala University, Box 516,
Se-751 20 Uppsala, Sweden}
\date{\today}
\begin{abstract}
The metric underlying the mixed state geometric phase in unitary and nonunitary evolution  
[Phys. Rev. Lett. {\bf 85}, 2845 (2000); Phys. Rev. Lett. {\bf 93}, 080405 (2004)] is delineated. 
An explicit form for the line element is derived and shown to be 
related to an averaged energy dispersion in the case of unitary evolution. The line element 
is measurable in interferometry involving nearby internal states. Explicit geodesics are found 
in the single qubit case. It is shown how the Bures line element can be obtained by 
extending our approach to arbitrary decompositions of density operators. The proposed 
metric is applied to a generic magnetic system in a thermal state. 
\end{abstract}
\maketitle
\section{Introduction} 
A quantum-mechanical metric underlies the notion of statistical 
distance that measures the distinguishability of quantum states \cite{wootters81,braunstein94}. 
Such measures can be used to quantify quantum entanglement \cite{shimony95,vedral97,wei03}, 
but have also found applications in the study of quantum phase transitions \cite{venuti07,you07}. 
Similarly, the related concept of path length has been used to find time-optimal curves in quantum 
state spaces \cite{carlini06} and to establish the speed-limit of quantum evolution 
\cite{margolus98,deffner13,mondal16}. 

Like the geometric phase (GP), the metric is closely related to the ray structure of quantum 
states. To each form of GP there is a corresponding metric. For pure states, the GP is the 
Aharonov-Anandan phase \cite{aharonov87} with the corresponding Fubini-Study metric 
\cite{provost80,anandan90}, both arising from the horizontal lift to the one-dimensional rays 
over the quantum state space. For mixed states, the GP can be taken as the Uhlmann holonomy 
\cite{uhlmann86} with the corresponding Bures metric \cite{hubner92} both arising from the 
horizontal lift to the possible decompositions of density operators. The horizontal lifts guarantee 
that the geometric quantities are properties of state space. 

The mixed state geometric phase (GP) in unitary \cite{sjoqvist00} and nonunitary 
\cite{tong04} evolution has been proposed as an alternative to Uhlmann's holonomy along 
paths of density operators. A key point of the mixed state GP is that it is operational in the 
sense that it is directly accessible in interferometry. Indeed, it has been studied on different 
experimental platforms \cite{du03,ericsson05,klepp08}. Although the mixed state GP is 
now a well-established concept in a wide range of contexts, the physics of the corresponding 
metric \cite{andersson19} has not been explored so far. The intention of 
the present work is to fill this gap. 

To understand the conceptual basis of our approach, we note that the corresponding 
mixed state GP in the case of unitary evolution reads \cite{sjoqvist00}
\begin{eqnarray}
\Phi_g = \arg \sum_{k} p_{k} e^{i\beta_{k}}
\label{eq:msgp}
\end{eqnarray}
with $p_{k}$ and $e^{i\beta_{k}}$ being eigenvalues and eigenstate GP factors, 
respectively, of the evolving density operator $\rho$. In other words, the spectral decomposition 
of $\rho$ plays a central role. Therefore, the corresponding metric must fundamentally 
be based on a distance for spectral decompositions of density operators. Here, we 
describe how such a metric can be designed. We further discuss various applications of this  
metric as well as its relation to the Bures' metric.  

\section{Derivation of line element} 
Consider a smooth path $t \mapsto \rho (t)$ of 
density operators representing the evolving state of a quantum system. We shall 
assume that all non-zero eigenvalues of $\rho (t)$ are non-degenerate. In this way, the 
gauge freedom in the spectral decomposition is the phase of the eigenvectors; thus, a 
non-degenerate density operator $\rho (t)$, assumed to have rank $N$, is in one to one 
correspondence with the $N$ orthogonal rays $\{ e^{if_{k} (t)} \ket{n_{k} (t)} | f_{k} (t) 
\in [0,2\pi) \}$. To capture this, we let 
\begin{eqnarray}
\mathcal{B} (t) = \left\{ 
\sqrt{p_{k} (t)} e^{if_{k} (t)} \ket{n_{k} (t)}
\right\}_{n=1}^N 
\end{eqnarray}
represent the spectral decompositions along the path. We further assume that all $f_{k} (t)$ 
are once differentiable.  

We propose the line element connecting two nearby points to be the minimum 
of the distance 
\begin{eqnarray}
 & & d^2 (t,t + dt) = 
\sum_{k} \left| \left| \sqrt{p_{k} (t)} e^{if_{k} (t)} \ket{n_{k} (t)} \right. \right.  
\nonumber \\ 
 & & \left. \left. - \sqrt{p_{k} (t+dt)} e^{if_{k} (t+dt)} \ket{n_{k} (t+dt)} \right|\right|^2  .  
\label{eq:distance}
\end{eqnarray}
To find this minimum, we expand the squares yielding 
\begin{eqnarray}
 & & d^2 (t,t + dt) 
 \nonumber \\ 
 & = & 2 - 2 \sum_{k} \sqrt{p_{k}(t) p_{k}(t+dt)} 
 \Big| \langle n_{k} (t) \ket{n_{k} (t+dt)} \Big| 
\nonumber \\ 
 & & \times \cos \lambda_k (t,t+dt) ,  
\label{eq:distance1}
\end{eqnarray}
where $\lambda_k (t,t+dt) = \dot{f}_k (t) dt + \arg \left[ 1 + \langle n_k (t) \ket{\dot{n}_k (t)} dt 
\right] + {\rm O} (dt^2)$. The line element is thus given by 
\begin{eqnarray}
& & ds^2 = d_{\min}^2 (t,t+dt) 
\nonumber \\ 
 & = & 2 - 2 \sum_{k} \sqrt{p_{k}(t) p_{k}(t+dt)} 
 \Big| \langle n_{k} (t) \ket{n_{k} (t+dt)} \Big| , 
\label{eq:metric}
\end{eqnarray}
being reached when all $\lambda_k (t,t+dt)$ vanish to first order in $dt$, which is equivalent to 
\begin{eqnarray}
\dot{f}_{k} (t) - i \langle n_{k} (t) \ket{\dot{n}_{k} (t)} = 0 , 
\label{eq:connection}
\end{eqnarray}
for all $k$. Equation (\ref{eq:connection}) is precisely the connection underlying the mixed 
state GP \cite{sjoqvist00}, itself a direct extension of the Aharonov-Anandan connection for 
pure states \cite{aharonov87}. The connection provides the necessary link between the 
mixed state GP \cite{sjoqvist00} and the metric concept considered here. 
Equation (\ref{eq:metric}) can be put on a more useful 
form by expanding to lowest non-trivial order in $dt$. We suppress the $t$ argument (for 
notational simplicity) and make use of the 
identities $\langle n_{k} \ket{\ddot{n}_{k}} + \langle \ddot{n}_{k} \ket{n_{k}} = 
- 2  \langle \dot{n}_{k} \ket{\dot{n}_{k}}$ and $\sum_{k} \dot{p}_{k} = 
\sum_{k} \ddot{p}_{k} = 0$, which follow from the normalization conditions 
$\langle n_{k} \ket{n_{k}} = 1$ and $\sum_{k} p_{k} = 1$. We find 
\begin{eqnarray}
ds^2 = \sum_{k} p_{k} ds_{k}^2 + \frac{1}{4} \sum_{k} \frac{dp_{k}^2}{p_{k}} , 
\label{eq:spectral_length}
\end{eqnarray}
where 
\begin{eqnarray}
ds_{k}^2 = \bra{\dot{n}_{k}} \left( \hat{1} - \ket{n_{k}} \bra{n_{k}} \right) 
\ket{\dot{n}_{k}} dt^2 
\end{eqnarray}
is the pure state Fubini-Study metric (infinitesimal line element) along $\ket{n_{k}}$ 
\cite{provost80} and $dp_{k} = \dot{p}_{k} dt$. Note the structural similarity between 
the first term of the right-hand side of Eq.~(\ref{eq:spectral_length}) and the expression for 
the mixed state GP in Eq.~(\ref{eq:msgp}), both being weighted sums of the corresponding 
pure state quantities. The second term we recognize as the Fischer-Rao 
information metric for classical probability distributions \cite{bengtsson06}. In the following, 
we shall examine various applications of the line element in Eq.~(\ref{eq:spectral_length}).  

\section{Applications}
\subsection{Unitary evolution, time-energy uncertainty} 
Let us first consider the case of unitary 
time evolution $i \hbar \dot{\rho} = [H,\rho]$ governed by some Hamiltonian $H$. Here, the 
Fischer-Rao term vanishes since the probability weights $p_{k}$ are constant. By using 
the geometric time-energy relation in Ref.~\cite{anandan90}, we find 
\begin{eqnarray}
ds^2 = \frac{1}{\hbar^2} \overline{\Delta E}^2 dt^2 , 
\label{eq:time_energy}
\end{eqnarray}
with the mixed state energy dispersion $\overline{\Delta E}^2 =  \sum_{k} p_{k} 
\left( \Delta_{k} E \right)^2$. Here, $\Delta_{k} E$ is the energy dispersion of 
$\ket{n_{k}}$. Thus, the speed by which the eigendecomposition of the density 
operator changes along the path is $ds/dt = (1/\hbar) \overline{\Delta E}$ . 

Note that the energy dispersion $\overline{\Delta E}^2$ is different from the standard 
quantum-mechanical dispersion $\Delta_{\rho} E^2 = {\rm Tr} (\rho H^2) - 
[{\rm Tr} (\rho H)]^2$. However, the inequality 
\begin{eqnarray}
\overline{\Delta E}^2 \leq \Delta_{\rho} E^2 
\label{eq:inequality}
\end{eqnarray}
relates the two. To prove this, we note that $\overline{\Delta E}^2$ and $\Delta_{\rho} E^2$ are 
independent of zero-point energy and are therefore unchanged under the shift $H \rightarrow 
\widetilde{H}  \equiv H - {\rm Tr} (\rho H)$.  We find $\Delta_{\rho} E^2 =  {\rm Tr} (\rho 
\widetilde{H}^2)$ and thus $\overline{\Delta E}^2 = {\rm Tr} (\rho \widetilde{H}^2) - 
\sum_{k} p_{k} \bra{n_{k}} \widetilde{H} \ket{n_{k}}^2 = \Delta_{\rho} E^2 - 
\sum_{k} p_{k} \bra{n_{k}} \widetilde{H} \ket{n_{k}}^2$, which implies 
Eq.~(\ref{eq:inequality}) since $\sum_{k} p_{k} \bra{n_{k}} \widetilde{H} 
\ket{n_{k}}^2 \geq 0$. 

A time-energy uncertainty relation similar to those of Refs.~\cite{anandan90,uhlmann92} 
can be formulated. Consider two unitarily connected states and assume $s_{\min}$ is 
the shortest distance between them, as measured by $ds$ in Eq.~(\ref{eq:spectral_length}). 
Let $\langle \overline{\Delta E} \rangle = (1/\Delta t) \int_0^{\Delta t} \overline{\Delta E} dt$ 
and $\langle \Delta_{\rho} E \rangle = (1/\Delta t) \int_0^{\Delta t} \Delta_{\rho} E dt$ be 
the time-averaged energy dispersions for the traversal time $\Delta t$ between the two 
states. Equation (\ref{eq:time_energy}) combined with Eq.~(\ref{eq:inequality}) implies 
\begin{eqnarray}
\langle \Delta_{\rho} E \rangle \Delta t \geq \langle \overline{\Delta E} \rangle \Delta t 
\geq s_{\min} \hbar , 
\end{eqnarray}
which provides a geometric lower bound for the energy-time uncertainty. This geometric  
bound is apparently tighter for $\langle \overline{\Delta E} \rangle$ than for 
$\langle \Delta_{\rho} E \rangle$. 

\subsection{Interferometry} 
We now address the operational significance of the line element 
$ds^2$. In the unitary case, the proposed line element can be related to measurable 
quantities by using the technique of Ref.~\cite{sjoqvist00}. Consider a Mach-Zehnder 
interferometer with a pair of 50-50 beam-splitters acting as $\ket{x} \mapsto 2^{-1/2} 
\big[ \ket{x} + (-1)^x \ket{x \oplus 1}\big]$ on the beam states $x=0,1$, and $\rho$ 
describing the `internal' state of the particles injected into the interferometer. 

\begin{figure}[htb]
\centering
\includegraphics[width=0.5\textwidth]{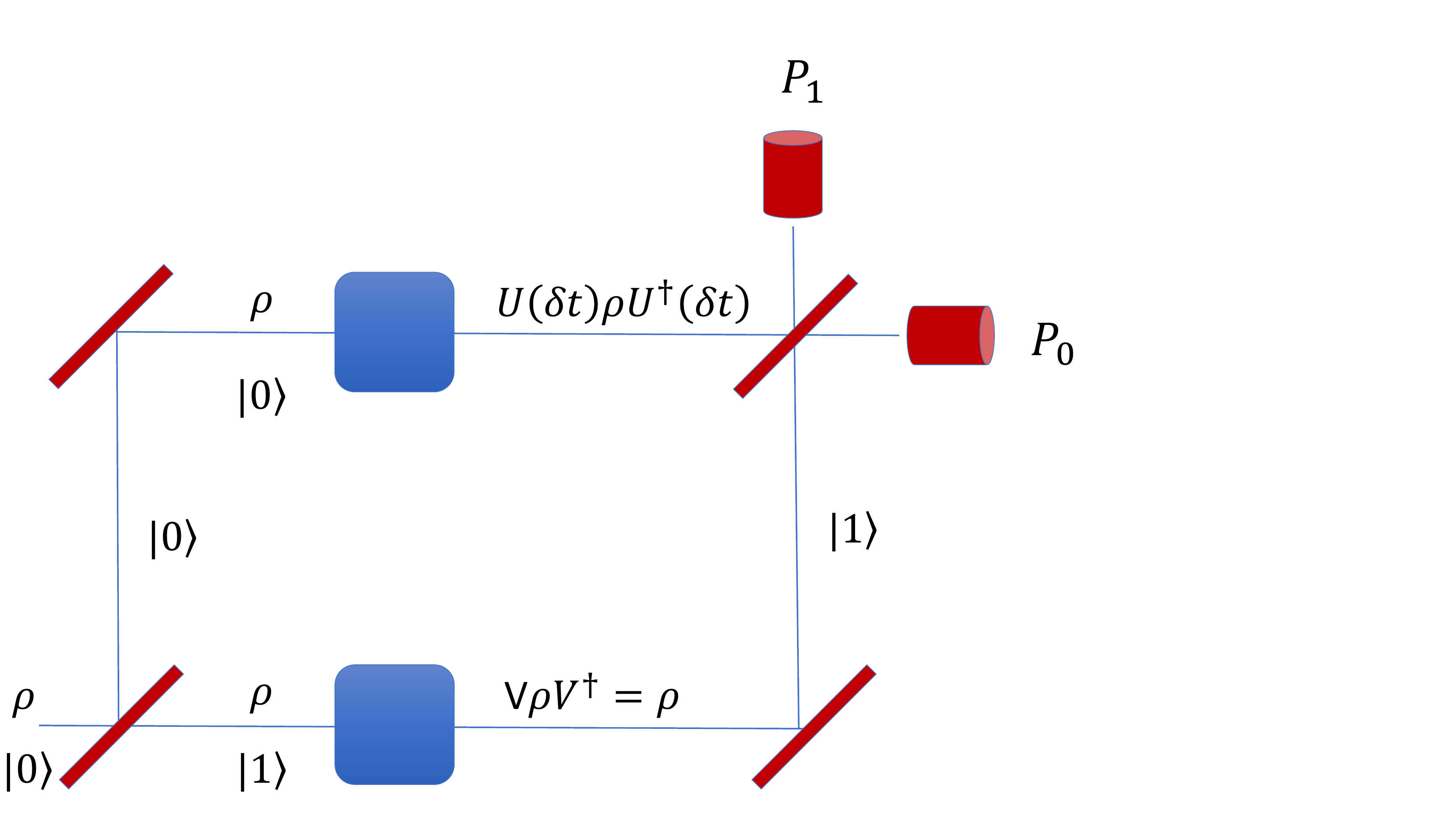}
\caption{Interferometer to measure the line element induced by the unitary $U(\delta t)$ 
in the $\ket{0}$ beam. The reference beam $\ket{1}$ is exposed to a unitary $V$ that 
commutes with the internal input state $\rho$. Its role is to maximize the output probability 
$P_0$ in the horizontal beam. To leading non-trivial order in $\delta t$, one has $P_0 = 
1 - \frac{1}{4} \delta s^2$, which gives direct experimental access to the line element for 
a small but finite time $\delta t$.}
\label{fig:interferometer}
\end{figure}

Assume the input state $\ket{0} \bra{0} \otimes \rho$ hits the first beam-splitter followed 
by a unitary $\ket{0} \bra{0} \otimes U(\delta t) + \ket{1} \bra{1} \otimes V$, $\delta t$ being 
a small but finite time interval and $[V,\rho]=0$. Thus, in the $0$-beam the internal state 
undergoes the transformation $\rho \mapsto U(\delta t) \rho U^{\dagger}(\delta t)$, while it 
remains unchanged in the $1$-beam: $\rho \mapsto V \rho V^{\dagger} = \rho$, see 
Fig.~\ref{fig:interferometer}. By writing $V = \sum_{k} e^{if_{k}} \ket{n_{k}} \bra{n_{k}}$, 
we obtain the probabilities 
\begin{eqnarray}
P_0 = 1 - P_1 = \frac{1}{2} + \frac{1}{2} {\rm Re} \sum_{k} p_{k} \bra{n_{k}} 
U(\delta t) \ket{n_{k}} e^{-if_{k}} 
\end{eqnarray}
to find the particles in the two beams after passing the second beam-splitter. We write 
$U(\delta t) = \hat{1} - \frac{i}{\hbar} H \delta t - \frac{1}{2\hbar^2} H^2 \delta t^2 + \ldots$, 
where $H$ is the Hamiltonian acting on the internal degrees of freedom of the particles, and 
maximize  $P_0$ over each of the phases $f_{k}$, yielding to lowest non-trivial order in 
$\delta t$ 
\begin{eqnarray}
P_{0,\max} = \max_{\{ f_{k} \}} P_0 = 1 - \frac{1}{4} \delta s^2 . 
\label{eq:prob}
\end{eqnarray}
Here, $\delta s^2 = \frac{1}{\hbar^2} \overline{\Delta E}^2 \delta t^2$
is Eq.~(\ref{eq:time_energy}) for a finite but small time interval. 

In order to generalize the interferometric setting to the non-unitary case, the  
purification-based technique described in Ref.~\cite{tong04} can be used.  
That is, one adds an auxiliary system and prepare the combined system in a 
pure internal input state $\ket{\Psi} = \sum_{k} \sqrt{p_{k}} \ket{n_{k}} \otimes 
\ket{a_{k}}$ with $\langle a_k \ket{a_l} = \delta_{kl}$, thus satisfying 
$\rho = {\rm Tr}_a \ket{\Psi} \bra{\Psi}$. Now, the above unitary 
that is applied between the beam-splitters is replaced by the extended unitary 
$\ket{0} \bra{0} \otimes W(\delta t) + \ket{1} \bra{1} \otimes V \otimes \hat{1}_a$. Here,  
$W(\delta t)$ acts on the combined system as $W(\delta t) \ket{\Psi} = 
\sum_{k} \sqrt{p_{k} + \delta p_{k}} U(\delta t) \ket{n_{k}} \otimes \ket{a_{k}}$, 
while $V \otimes \hat{1}_a \ket{\Psi} = \sum_{k} e^{if_{k}} \sqrt{p_{k}} \ket{n_{k}} 
\otimes \ket{a_{k}}$. The reduced states in the two beams undergo 
the transformations $\rho \mapsto U(\delta t) \sum_{k} (p_{k} + \delta p_{k}) 
\ket{n_{k}} \bra{n_{k}} U^{\dagger} (\delta t)$ and $\rho \mapsto V \rho V^{\dagger} = 
\rho$. By superposing the two beams at the second beam-splitter, we obtain 
the output state 
\begin{eqnarray}
\ket{\Psi_{\rm out}} & \propto &  
\sum_{k} \left( \sqrt{p_{k} + \delta p_{k}} U(\delta t) \ket{n_{k}} \right. 
\nonumber \\ 
 & & \left. + e^{if_{k}} \sqrt{p_{k}} \ket{n_{k}} \right) \otimes \ket{a_{k}} , 
\end{eqnarray}
which results in the probability in Eq.~(\ref{eq:prob}) with the Fischer-Rao-like term 
$\frac{1}{4} \sum_{k} \delta p_{k}^2/p_{k}$ being added to $\delta s^2$. Compared 
to the above unitary interferometric setting, the non-unitary scheme is clearly more 
demanding as it would require a substantially higher level of control of interacting 
quantum systems. 

\subsection{Qubit geodesics} 
Geodesics contain important information about the curved space 
that  is described by the metric. Here, we demonstrate that the geodesics associated with 
$ds$ in Eq.~~(\ref{eq:spectral_length}) and connecting arbitrary non-degenerate ($r\neq 0$) 
states of a single qubit can be found analytically. 

\begin{figure}[htb]
\centering
\includegraphics[width=0.36\textwidth]{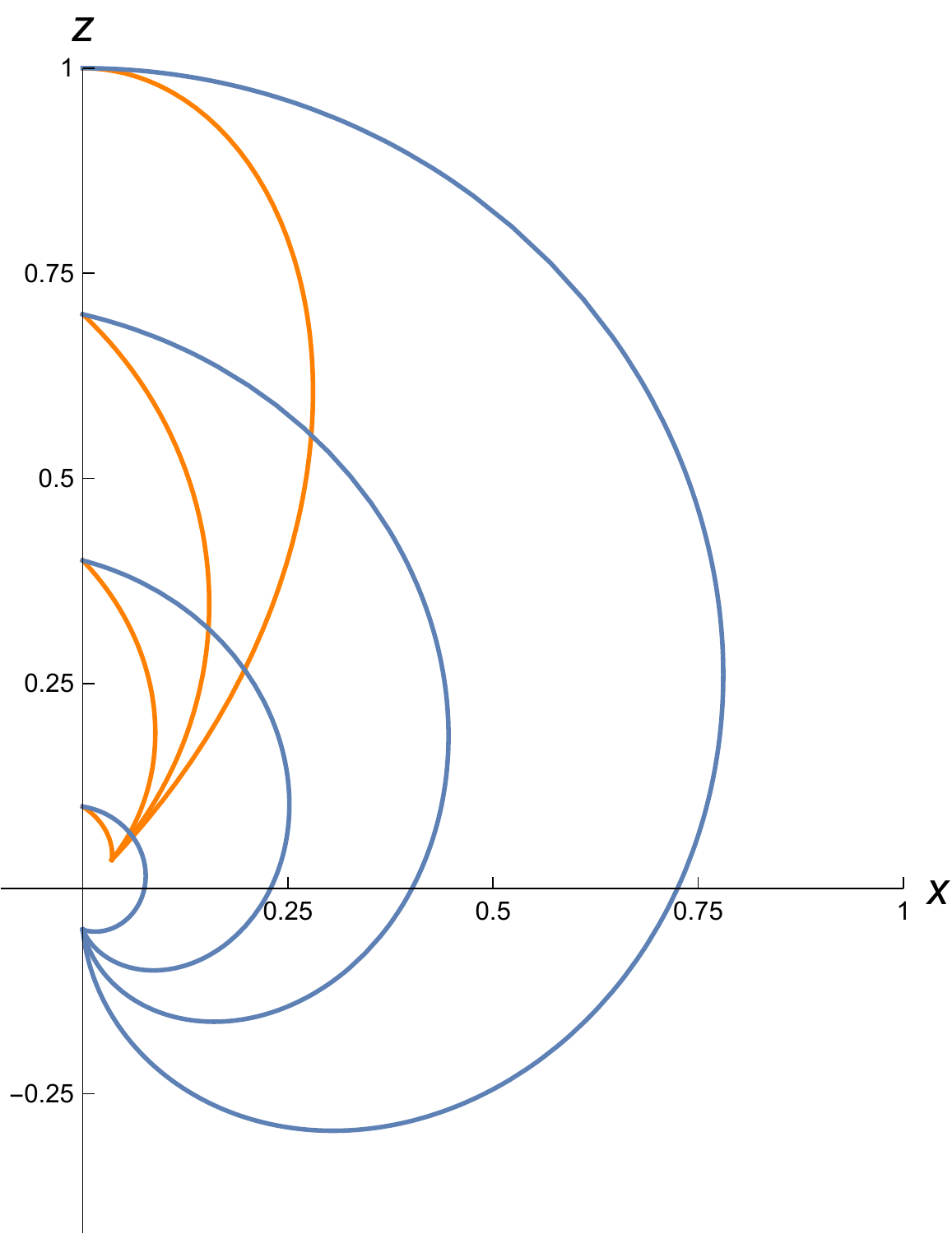}
\caption{Single-qubit geodesic curves in the $xz$-plane of the Bloch ball starting and ending 
at polar coordinates $(r_1,0)$ and $(r_2,\theta_{12})$, respectively. The curves have the form 
$r(\theta) (\sin \theta,0,\cos\theta)$ with $r(\theta)$ given by Eq.~(\ref{eq:qubit_geo}). We 
have chosen $r_1 = r(0) = 0.1, 0.4,0.7,1$ and $r_2 = r(\theta_{12}) = 0.05$. The angular 
position of the end-points are $\theta_{12} = \frac{\pi}{4}$ (orange curves) and 
$\theta_{12} = \pi$ (blue curves).}
\label{fig:geodesics}
\end{figure}

First note that $ds_0^2 = ds_1^2 = \frac{1}{4} \left( d\theta^2 + \sin^2 \theta d\phi^2 \right) 
\equiv \frac{1}{4} ds_{\mathcal{S}^2}^2$ with $\theta$ and $\phi$ the polar angles on 
the Bloch sphere. We further write $p_0 = 1 - p_1 = \frac{1}{2} (1+r)$, $r \neq 0$, in terms 
of which Eq.~(\ref{eq:spectral_length}) takes the form \cite{remark}
\begin{eqnarray}
ds^2 =  \frac{1}{4} \left( \frac{dr^2}{1-r^2} + ds_{\mathcal{S}^2}^2 \right) . 
\label{eq:qubit_line_element}
\end{eqnarray}
The geodesics are found by minimizing $\int ds$ over all curves connecting pairs of  
points in the Bloch ball. The curve that provides the minimum for a given pair must lie 
in a plane that contains the origin of the Bloch ball. By choosing the $xz$-plane ($\phi =0$), 
we look for a curve that connects points at polar coordinates $(r_1,0)$ and $(r_2,\theta_{12})$. 
We thus wish to find the curve $\theta \in [0,\theta_{12}] \mapsto 
r_{\rm g}(\theta)$ that minimizes the length 
\begin{eqnarray}
l(\theta_{12}) & = & \frac{1}{2} \int_0^{\theta_{12}} \sqrt{1+\frac{r'^2}{1-r^2}} d\theta 
\nonumber \\ 
 & = & \frac{1}{2} \int_0^{\theta_{12}} \mathcal{L}(r,r') d\theta, 
\label{eq:qubit_le}
\end{eqnarray}
where we use the short-hand notation $r' = \frac{d}{d\theta} r(\theta)$ and $r=r(\theta)$. 
The Euler-Lagrange equation can be solved by means of Beltrami's identity 
\begin{eqnarray}
\frac{\partial \mathcal{L}}{\partial r'} r' - \mathcal{L} = c,
\end{eqnarray}
the constant $c$ being determined by the boundary conditions $r(0) = r_1$ and 
$r(\theta_{12}) = r_2$. We find 
\begin{eqnarray}
r_{\rm g} (\theta) = \sin \left[ \arcsin r_1 + \left( \arcsin r_2 - \arcsin r_1 \right) 
\frac{\theta}{\theta_{12}} \right] . 
\label{eq:qubit_geo}
\end{eqnarray}
Figure \ref{fig:geodesics} shows some examples of geodesic curves in the Bloch ball. 

The length of the geodesics can be computed by inserting Eq.~(\ref{eq:qubit_geo}) into 
Eq.~(\ref{eq:qubit_le}) and performing the integration. One finds 
\begin{eqnarray}
l_{\rm g} = \frac{1}{2} \sqrt{\theta_{12}^2 + \left( \arcsin r_2 - \arcsin r_1 \right)^2} . 
\end{eqnarray}
We note that the geodesics for $r_2=r_1$ are circle arcs of length $\theta_{12}/2$, which 
is half the geodesic distance on $\mathcal{S}^2$. For pure ($r_1=r_2=1$) states, this is 
consistent with the Fubini-Study distance for single qubits \cite{anandan90}. $l_g$ measures 
the distance between non-degenerate qubit states. 

\subsection{Thermal magnetic systems} 
We illustrate the metric in Eq.~(\ref{eq:spectral_length}) by considering the response of a 
magnetic system in a thermal state to changes in temperature $T$ and in an applied 
magnetic field $b$. This is modeled by the Hamiltonian $H(b) = H_0 + b S_z$, $H_0$ 
being a generic Hamiltonian describing interactions between a collection of spins and 
$S_z$ is the total spin of the system. Let $\{ \ket{m(b)} \}$ and $\{ \varepsilon_m (b) \}$ 
be eigenstates and eigenvalues, respectively, of $H(b)$. The thermal state takes the form 
$\rho = e^{-\beta H(b)}/Z$ with $Z = {\rm Tr} \left( e^{-\beta H(b)} \right)$ the partition 
function and $\beta$ the inverse temperature. The following analysis shows that the 
metric can be related to thermodynamic quantities. 

Let us first consider changes in temperature. One finds 
\begin{eqnarray}
ds^2 & = & \frac{C_V}{4\beta^2} d\beta^2 ,  
\end{eqnarray}
where $C_V$ is specific heat for a Boltzmann distribution, being related to the energy 
fluctuations according to $C_V = \beta^2 \left( \langle \varepsilon^2 \rangle - 
\langle \varepsilon \rangle^2 \right)$. Here and in the following, $\langle \cdot\rangle$ 
is the thermodynamic average obtained by means of the Boltzmann factors 
$p_m = e^{-\beta \varepsilon_m (b)}/Z$. For changes in the applied magnetic field, 
we find 
\begin{eqnarray}
ds^2 & = & \left( \frac{\beta \chi_M}{4} + \sum_m p_m \chi_{F,m} \right) db^2   
\end{eqnarray}
with the magnetic susceptibility $\chi_M = \beta \left[ \left\langle \left( \partial \varepsilon / 
\partial b \right)^2 \right\rangle - \left\langle \partial \varepsilon / \partial b \right\rangle^2 \right]$ 
and the fidelity susceptibility 
\cite{you07} 
\begin{eqnarray}
 \chi_{F,m} (b) = \sum_{m' \neq m} \frac{\left| \bra{m'} S_z \ket{m} \right|^2}{(\varepsilon_m - 
\varepsilon_{m'})^2}
\end{eqnarray}
of state $m$. 

\section{Relation to Bures' metric} 
Before concluding, we justify our distance concept by  
demonstrating that the Bures metric \cite{hubner92,uhlmann92} can be obtained if we 
extend Eq.~(\ref{eq:distance}) to arbitrary decompositions of $\rho (t)$. We use that the set 
\begin{eqnarray}
\mathcal{A} (t) & = & \left\{ \sum_{l} \sqrt{p_{l} (t)} 
\ket{n_{l} (t)} \mathbb{V}_{lk} (t) \right\} 
\end{eqnarray}
of sub-normalized vectors is a decomposition of $\rho (t)$ for any unitary $N' \times N'$ 
matrix $\mathbb{V}$ with $N'-N$ zero vectors added \cite{schrodinger36,hughston93}. If  
$\mathcal{B} (t)$ is replaced by $\mathcal{A} (t)$ in Eq.~(\ref{eq:distance}), we find the 
distance 
\begin{eqnarray}
\tilde{d}^2 (t,t+dt) = 2 - 2 {\rm Re} {\rm Tr} \left[ \mathbb{M}_t (dt) \mathbb{V} (t+dt)  
\mathbb{V}^{\dagger} (t) \right] , 
\end{eqnarray}
where $\left[ \mathbb{M}_{t} (dt) \right]_{kl} = 
\sqrt{p_{k} (t) p_{l} (t+dt)} \langle n_{k} (t) \ket{n_{l} (t+dt)}$ is the overlap matrix. By 
means of the polar decomposition $\mathbb{M}_t (dt) = \big| \mathbb{M}_t (dt) \big| 
\mathbb{U}_t (dt)$, we find the line element 
\begin{eqnarray}
d\tilde{s}^2 = \tilde{d}_{\min}^2 (t,t+dt) = 2 - 2 {\rm Tr} \big| \mathbb{M}_t (dt) \big| , 
\label{eq:minimaldistance}
\end{eqnarray}
by choosing $\mathbb{V}$ such that $\mathbb{U}_t (dt) \mathbb{V} (t+dt)  
\mathbb{V}^{\dagger} (t) = \mathbb{I}$. One may use the spectral form of $\rho (t)$ 
and $\rho (t+dt)$ and the orthonormality of $\left\{ \ket{n_{k} (t)} \right\}$ to obtain  
\begin{eqnarray}
 & & \sqrt{\rho (t)} \rho (t+dt) \sqrt{\rho (t)} 
\nonumber \\ 
 & = & \left( \sum_{k,k'} \ket{n_{k} (t)} \left| \mathbb{M}_t (dt) \right|_{kk'}  
\bra{n_{k'} (t)} \right) 
\nonumber \\  
 & & \times \left( \sum_{l',l} \ket{n_{l'} (t)} \left| \mathbb{M}_t (dt) \right|_{l' l} 
\bra{n_{l} (t)} \right) , 
\label{eq:squared_fidelity}
\end{eqnarray}
from which we conclude 
\begin{eqnarray}
 & & \sqrt{\sqrt{\rho (t)} \rho (t+dt) \sqrt{\rho (t)}} 
 \nonumber \\ 
  & = & \sum_{k,l} \ket{n_{k} (t)} \left| \mathbb{M}_t (dt) \right|_{kl}  \bra{n_{l} (t)} . 
\label{eq:operator_fidelity}
\end{eqnarray}
By taking the trace, we see that Eq.~(\ref{eq:minimaldistance}) can be expressed as 
\begin{eqnarray}
d\tilde{s}^2 = 
2 - 2 {\rm Tr} \sqrt{\sqrt{\rho (t)} \rho (t+dt) \sqrt{\rho (t)}} , 
\label{eq:uhlmannbures}
\end{eqnarray}
which is precisely the Bures line element \cite{hubner92,uhlmann92}.  

\section{Conclusions} 
The concept of metric associated with the spectral decomposition of 
mixed quantum states is delineated and its physical significance discussed. This completes 
the theory of mixed state GP proposed in Ref.~\cite{sjoqvist00}, in the same way as the 
Fubini-Study and the Bures metric complete the theory of pure state geometric phase and 
Uhlmann holonomy, respectively. The relation to energy-time uncertainty and thermodynamics 
found above, suggests that the proposed metric can be expected to find applications in the 
problem of finding time-optimal evolutions of mixed quantum states as well as 
in the study of phase transitions in many-body quantum systems at non-zero temperatures. 

\section*{Acknowledgments} 
This work was supported by the Swedish Research Council (VR) through Grant No. 
2017-03832.


\begin{thebibliography}{99}
\bibitem{wootters81} W. K. Wootters, 
Statistical distance and Hilbert space, 
Phys. Rev. D {\bf 23}, 357 (1981).
\bibitem{braunstein94} S. L. Braunstein and C. M. Caves, 
Statistical distance and the geometry of quantum states, 
Phys. Rev. Lett. {\bf 72}, 3439 (1994).  
\bibitem{shimony95} A. Shimony, 
Degree of Entanglement, 
Ann. N.Y. Acad. Sci. {\bf 755}, 675 (1995).
\bibitem{vedral97} V. Vedral, M. B. Plenio, M. A. Rippin, and P. L. Knight, 
Quantifying Entanglement, 
Phys. Rev. Lett. {\bf 78}, 2275 (1997). 
\bibitem{wei03} T.-C. Wei and P. M. Goldbart, 
Geometric measure of entanglement and applications to bipartite and multipartite 
quantum states, 
Phys. Rev. A {\bf 68}, 042307 (2003). 
\bibitem{venuti07} L. C. Venuti and P. Zanardi, 
Quantum Critical Scaling of the Geometric Tensors, 
Phys. Rev. Lett. {\bf 99}, 095701 (2007).  
\bibitem{you07} W.-L. You, Y.-W. Li, S.-J. Gu, 
Fidelity, dynamic structure factor, and susceptibility in critical phenomena, 
Phys. Rev. E {\bf 76}, 022101 (2007). 
\bibitem{carlini06} A. Carlini, A. Hosoya, T. Koike, and Y. Okudaira, 
Time-Optimal Quantum Evolution, 
Phys. Rev. Lett. {\bf 96}, 060503 (2006). 
\bibitem{margolus98} N. Margolus, L. B. Levitin, 
The maximum speed of dynamical evolution, 
Physica D {\bf 120}, 188 (1998). 
\bibitem{deffner13} S. Deffner and E. Lutz, 
Quantum Speed Limit for Non-Markovian Dynamics, 
Phys. Rev. Lett. {\bf 111}, 010402 (2013). 
\bibitem{mondal16} D. Mondal and A. K. Pati, 
Quantum speed limit for mixed states using an experimentally realizable metric,
Phys. Lett. A {\bf 380}, 1395 (2016).  
\bibitem{aharonov87} Y. Aharonov and J. Anandan, 
Phase change during a cyclic quantum evolution, 
Phys. Rev. Lett. {\bf 58}, 1593 (1987). 
\bibitem{provost80} J. P. Provost and G. Vallee, 
Riemannian structure on manifolds of quantum states,
Commun. Math. Phys. {\bf 76}, 289 (1980). 
\bibitem{anandan90} J. Anandan and Y. Aharonov, 
Geometry of Quantum Evolution, 
Phys. Rev. Lett. {\bf 65}, 1697 (1990). 
\bibitem{uhlmann86} A. Uhlmann, 
Parallel transport and quantum holonomy along density operators, 
Rep. Math. Phys. {\bf 24}, 229 (1986). 
\bibitem{hubner92} M. H\"ubner, 
Explicit computation of the Bures distance for density matrices, 
Phys. Lett. A {\bf 163}, 239 (1992).  
\bibitem{sjoqvist00} E. Sj\"{o}qvist, A. K. Pati, A. Ekert, J. S. Anandan, M. Ericsson,
D. K. L. Oi, and V. Vedral, 
Geometric Phases for Mixed States in Interferometry, 
Phys. Rev. Lett. {\bf 85}, 2845 (2000). 
\bibitem{tong04} D. M. Tong, E. Sj\"{o}qvist, L. C. Kwek, and C. H. Oh, 
Kinematic Approach to the Mixed State Geometric Phase in Nonunitary Evolution, 
Phys. Rev. Lett. {\bf 93}, 080405 (2004). 
\bibitem{du03} J. Du, P. Zou, M. Shi, L. C. Kwek, J.-W. Pan, C. H. Oh, A. Ekert, D. K. L. Oi, 
and M. Ericsson, 
Observation of Geometric Phases for Mixed States using NMR Interferometry, 
Phys. Rev. Lett. {\bf 91}, 100403 (2003); 
\bibitem{ericsson05} M. Ericsson, D. Achilles, J. T. Barreiro, D. Branning, N. A. Peters, 
and P. G. Kwiat, 
Measurement of Geometric Phase for Mixed States Using Single Photon Interferometry, 
Phys. Rev. Lett. {\bf 94}, 050401 (2005). 
\bibitem{klepp08} J. Klepp, S. Sponar, S. Filipp, M. Lettner, G. Badurek, and Y. Hasegawa, 
Observation of Nonadditive Mixed-State Phases with Polarized Neutrons, 
Phys. Rev. Lett. {\bf 101}, 150404 (2008). 
\bibitem{andersson19} O. Andersson, 
Holonomy in Quantum Information Geometry, 
arxiv:1910.08140. 
\bibitem{uhlmann92} A. Uhlmann, 
An energy dispersion estimate,  
Phys. Lett. A {\bf 161}, 329 (1992). 
\bibitem{remark} The absence of the $r^2$ factor in the angular part of the line element 
implies that the metric is singular at the origin of the Bloch ball, corresponding to a 
degenerate qubit state. This is an illustration of the general fact that the metric is undefined 
for degenerate density operators. 
\bibitem{schrodinger36} E. Schr\"odinger, 
Probability relations between separated systems, 
Proc. Camb. Phil. Soc. {\bf 32}, 446 (1936). 
\bibitem{hughston93} L. P. Hughston, R. Jozsa, and W. K. Wootters, 
A complete classification of quantum ensembles having a given density matrix, 
Phys. Lett. A {\bf 183}, 14 (1993).
\bibitem{bengtsson06} I. Bengtsson and K. \.{Z}yczkowski,   
{\it Geometry of quantum states} (Cambridge University Press,
Cambridge, 2006). 
\end{thebibliography}
\end{document}